\documentclass[journal]{IEEEtran}
\usepackage{graphicx,amssymb,amsmath}
\usepackage{multicol}
\usepackage{CJK}
\usepackage{amsmath}
\usepackage[noadjust]{cite}
\usepackage{setspace}
\usepackage{stfloats}
\usepackage{midfloat}
\usepackage{flushend,cuted}
\usepackage{bm}
\usepackage{multirow}
\usepackage{array}
\usepackage{stfloats}
\usepackage{cite}
\usepackage{subfig}

\ifCLASSINFOpdf
\else
\fi
\newcommand{\tr}{\textmd{tr}}
\newcommand{\E}{\textmd{E}}
\newcommand{\diag}{\textmd{diag}}

\newcommand{\vect}{\textmd{vec}}

\newcounter{MYtempeqncnt}
\begin{document}
%
\title{Asymptotic Analysis for Low-Resolution Massive MIMO Systems with MMSE Receiver}

\author{Kai Liu,
        Cheng Tao,
        Liu Liu
        and~Yinsheng Liu
\thanks{K. Liu, C. Tao, L. Liu are with the Institute of Broadband
Wireless Mobile Communications, Beijing Jiaotong University, Beijing, 100044, China (email:
liukai@bjtu.edu.cn; chtao@bjtu.edu.cn; liuliu@bjtu.edu.cn).}
\thanks{Y. Liu is with the State Key Laboratory   of Rail Traffic Control and Safety, Beijing Jiaotong University, Beijing, 100044, China (email: ys.liu@bjtu.edu.cn).}
}


%


\maketitle

\begin{abstract}
The uplink achievable rate of massive multiple-input-multiple-output (MIMO) systems, where the low-resolution analog-to-digital converters (ADCs) are assumed to equip at the base station (BS), is investigated in this paper. We assume that only imperfect channel station information is known at the BS. Then a new MMSE receiver is designed by taking not only the Gaussian noise, but also the channel estimation error and quantizer noise into account. By using the Stieltjes transform of random matrix, we further derive a tight asymptotic equivalent for the uplink achievable rate with proposed MMSE receiver. We present a detailed analysis for the number of BS antennas through the expression of the achievable rates and validate the results using numerical simulations. It is also shown that we can compensate the performance loss due to the low-resolution quantization by increasing the number of antennas at the BS.
\end{abstract}



%
\IEEEpeerreviewmaketitle

\section{Introduction}
Massive multiple-input multiple-output (MIMO), as well-known as one of the key techniques for 5G wireless communication, has attracted a lot of research interest currently. The main idea of massive MIMO is to equip hundreds of antennas at the base station such that simple detection processing or transmit precoding can be employed in order to reduce  the effect of noise and user interference, and hence, the spectral efficiency can be significantly improved \cite{marzetta2010noncooperative,ngo2011energy,rusek2013scaling,lu2014an}.

During the process of the massive MIMO deployment, howerver,  perhaps the most important issue that hinders its commercialization is the system power consumption and economic cost. Since massive MIMO systems have hundreds of BS antennas, the system power consumption will be prohibited if each BS antenna equips with power-hungered ADCs. Therefore, the research interest has shiffted to reducing the power consumption of ADCs recently. In doing so, employing the low-resolution ADCs becomes a potential solution \cite{chiara2014massive,jianhua2014high,wang2015multiuser,yongzhi2017channel}.

By considering the exact nonlinear property of a quantizer, the authors in \cite{singh2009on} first evaluated the communication transmission limits imposed by low-resolution ADCs. However, it is difficult to deal with a nonlinear operation in general. Then in order to simplify the analysis, \cite{fletcher2007robust} proposed an additive quantization noise model (AQNM), which approximately reformulates the nonlinear quantizer as a linear one by treating the quantizer noise as additive and independent noise. By employing AQNM, there has been many work on the performance analysis for the low-resolution massive MIMO systems \cite{bai2013on,orhan2015low,fan2015uplink,zhang2017performance,jacobasson2017throughput}. \cite{orhan2015low} investigated the effect of ADC resolution and bandwidth on the achievable rate for a multi-antenna system using the AQNM. \cite{fan2015uplink} derived an approximation of the uplink achievable rate for low-resolution massive MIMO systems with MRC receiver over Rayleigh fading channels. Then the result to the Rician fading channels has been extended in \cite{zhang2017performance}. However, most of these prior work simply assumed that perfect channel state information (CSI) is known at the BS, which is not well justified in practical systems.

In this paper, the asymptotic uplink achievable rate for low-resolution massive MIMO systems is investigated. More precisely, we assume that imperfect CSI, which is acquired by pilot training, is known at the BS. Then, in contrast to prior work, we properly design the MMSE receiver that takes not only the AWGN but also the channel estimation error and quantization noise into account. Employing the proposed MMSE receiver, we then derive an asymptotic equivalent for the uplink achievable rate using the Stieltjes transform of random matrix. Finally, we present numerical results to verify our theoretical analysis. 


\section{System Model}
In this section, a $K$ single-antenna users and $M$ BS-antenna massive MIMO system, where each antenna is assumed to be equipped with low-resolution ADCs, is considered. Since the uplink transmission is considered in this paper, we assume that the data transmitted from all $K$ users to BS is independent, thus the analog signal received at the BS is
\begin{equation}
  \mathbf{y} = \sqrt{p_{\rm u}}\mathbf{G}\mathbf{x} + \mathbf{n},
\end{equation}
where $\mathbf{G}=\mathbf{H}\mathbf{D}^{1/2}\in\mathbb{C}^{M\times K}$ is the channel matrix, $\mathbf{n}\sim\mathcal{CN}(\mathbf{0},\mathbf{I})\in\mathbb{C}^{M\times 1}$ is the AWGN, and $\mathbf{x}\in\mathbb{C}^{K\times 1}$ is the data symbol vector, which satisfies the assumption of $\E\{\mathbf{x}\mathbf{x}^H\} = \mathbf{I}$. More precisely, $\mathbf{D}\in\mathbb{C}^{K\times K}$ is a diagonal matrix with the $k$th diagonal term $d_k$ being the large-scale fading coefficient of the $k$th user, and $\mathbf{H}\in\mathbb{C}^{M\times K}$ denotes the small-scale fading channel matrix. In order to facilitate the representation, we vectorize the channel matrix $\mathbf{G}$ as $\underline{\mathbf{g}} = \vect(\mathbf{G})$, and hence we have $\underline{\mathbf{g}}\sim\mathcal{CN}(0, \mathbf{\Sigma}_{\underline{\mathbf{g}}})$.

In this paper, the Rayleigh fading is assumed and hence, 
according to the definition of channel matrix $\mathbf{G}$, we can readily obtain the covariance matrix $\mathbf{\Sigma}_{\underline{\mathbf{g}}}$ as
\begin{equation}\label{C_g}
  \mathbf{\Sigma}_{\underline{\mathbf{g}}} = \mathbf{D}\otimes\mathbf{I}.
\end{equation}

After the low-resolution ADCs, the quantized digital signal obtained  can be  represented as
\begin{equation}\label{quantized_sig}
  \mathbf{z}_{\rm u} = \mathcal{Q}(\mathbf{y}) = \mathcal{Q}(\sqrt{p_{\rm u}}\mathbf{G}\mathbf{x} + \mathbf{n}_{\rm u}),
\end{equation}
where $\mathcal{Q}(.)$ is the quantizer function. As we can see from \eqref{quantized_sig} that the quantized signal is complicated to deal with due to the nonlinear quantizer function. However, it is shown in \cite{fletcher2007robust} that if the gain of the automatic gain control is set appropriately and quantizer input is distributed as Gaussian, then the additive quantizer noise model (AQNM) can be employed to reformulate the quantized signal as
\begin{align}
  \mathbf{z}_{\rm u} &= \kappa\mathbf{y} + \mathbf{q}_{\rm u}  \nonumber \\
  &= \kappa\sqrt{p_{\rm u}}\mathbf{G}\mathbf{x} + \kappa\mathbf{n}_{\rm u}+ \mathbf{q}_{\rm u},
\end{align}
where $\kappa = 1-\alpha$ and $\alpha$ is the inverse of the signal-to quantization-noise ratio \cite{fletcher2007robust}. 
More precisely, \cite{fan2015uplink} listed the values of $\alpha$ for ADCs with 1-3 quantization bits. $\mathbf{q}_{\rm u}\sim\mathcal{CN}(\mathbf{0},\mathbf{\Sigma}_{\mathbf{q}_{\rm u}})\in\mathbb{C}^{M\times 1}$ is the additive Gaussian quantizer noise that is independent with $\mathbf{y}$. According to \cite{fan2015uplink}, the covariance matrix $\mathbf{\Sigma}_{\mathbf{q}_{\rm u}}$ of quantizer noise $\mathbf{q}_{\rm u}$ for individual channel realization is given by
\begin{equation}
  \mathbf{\Sigma}_{\mathbf{q}_{\rm u}} = \alpha\kappa\diag(p_{\rm u}\mathbf{G}\mathbf{G}^H + \mathbf{I}).
\end{equation}

\section{Channel Estimation and Data Transmission}
There has been limited prior work on the achievable rate analysis for the low-resolution massive MIMO systems with MRC/ZF receiver \cite{fan2015uplink,zhang2017performance,jacobasson2017throughput}. However, these prior work only assumed an ideal case that perfect CSI is known at the BS.

In contrast to these prior work, we assume that the CSI needs to be estimated at the BS by transmitting pilot sequences. Owing to the nonlinear quantization function, the channel estimation problem is challenging in low-resolution massive MIMO systems. Up to now, there has been many literatures working on this problem and different algorithms are proposed \cite{juncil2015near,yongzhi2017channel}. Inspired by the recent work in \cite{yongzhi2017channel}, we will employ the LMMSE method to estimate the channels in the following.

\subsection{Channel Estimation Stage}

In the channel estimation stage, $K$ users are assumed to transmit orthogonal pilot sequences  to the BS simultaneously, i.e., $\mathbf{F}^H\mathbf{F} = \tau \mathbf{I}_K$, where $\mathbf{F}\in\mathbb{C}^{\tau\times K}$ represents the pilot matrix and $\tau$ is the length of the pilot.  Then according to the AQNM, the quantized pilot signal matrix can be expressed as
\begin{align}\label{R_p}
  \mathbf{Z}_{\rm p} &= \mathcal{Q}(\sqrt{p_{\rm p}}\mathbf{G}\mathbf{F}^T + \mathbf{N}_{\rm p})\nonumber \\
  & = \kappa\sqrt{p_{\rm p}}\mathbf{G}\mathbf{F}^T +  \kappa\mathbf{N}_{\rm p} + \mathbf{Q}_{\rm p}
\end{align}
where $p_{\rm p}$ is the pilot transmit power, $\mathbf{N}_{\rm p}\in\mathbb{C}^{M\times \tau}$ is the AWGN matrix. $\mathbf{Q}_{\rm p}\in\mathbb{C}^{M\times \tau}$ is the quantizer noise in the channel estimate stage.

After vectorizing $\mathbf{Z}_{\rm p}$ as $\mathbf{z}_{\rm p}$, we have
\begin{equation}\label{r_p}
  \mathbf{z}_{\rm p} = \kappa\sqrt{p_{\rm p}}\bar{\mathbf{F}}\underline{\mathbf{g}} +  \kappa\mathbf{n}_{\rm p} + \mathbf{q}_{\rm p},
\end{equation}
where $\bar{\mathbf{F}} = \mathbf{F}\otimes \mathbf{I}$, $\mathbf{n}_{\rm p }=\vect(\mathbf{N}_{\rm p})$ and $\mathbf{q}_{\rm p }=\vect(\mathbf{Q}_{\rm p})$. Accordingly, the covariance matrix of $\mathbf{q}_{\rm p}$ can be given by
\begin{equation}
  \mathbf{\Sigma}_{\mathbf{q}_{\rm p}} = \alpha\kappa\diag\left(p_{\rm p}\bar{\mathbf{F}}\mathbf{\Sigma}_{\underline{\mathbf{g}}}\mathbf{F}^H + \mathbf{I}\right).
\end{equation}

By following the standard reasoning of LMMSE method, then the LMMSE channel estimate can be given by
\begin{equation}\label{hat_g}
    \hat{\underline{\mathbf{g}}} = \kappa\sqrt{p_{\rm p}}\mathbf{\Sigma}_{\underline{\mathbf{g}}}\bar{\mathbf{F}}^H\mathbf{\Sigma}_{\mathbf{z}_{\rm p}}^{-1}\mathbf{z}_{\rm p},
\end{equation}
where
\begin{equation}\label{C_r_p}
  \mathbf{\Sigma}_{\mathbf{z}_{\rm p}} = \kappa^2 p_{\rm p}\bar{\mathbf{F}} \mathbf{\Sigma}_{\underline{\mathbf{g}}} \bar{\mathbf{F}}^H + \kappa^2\mathbf{I} + \mathbf{\Sigma}_{\mathbf{q}_{\rm p}}.
\end{equation}

Considering \eqref{hat_g}, we can further obtain the covariance matrix of $\hat{\underline{\mathbf{g}}}$ as
\begin{equation}\label{C_hat_g}
  \mathbf{\Sigma}_{\hat{\underline{\mathbf{g}}}} = \kappa^2 p_{\rm p} \mathbf{\Sigma}_{\underline{\mathbf{g}}}\bar{\mathbf{F}}^H\mathbf{\Sigma}_{\mathbf{z}_{\rm p}}^{-1}\bar{\mathbf{F}}\mathbf{\Sigma}_{\underline{\mathbf{g}}}.
\end{equation}

Inspired by \cite{yongzhi2017channel}, we  consider a discrete Fourier transform (DFT) matrix as the pilot matrix in order to derive a concise expression of the covariance matrix $\mathbf{\Sigma}_{\hat{\underline{\mathbf{g}}}}$. Therefore, by noting $\mathbf{\Sigma}_{\underline{\mathbf{g}}} = \mathbf{D}\otimes \mathbf{I},$ we can obtain the covariance matrix  $\mathbf{\Sigma}_{\mathbf{q}_{\rm p}}$ of the quantizer noise $\mathbf{q}_{\rm p}$ as
\begin{equation}\label{C_q_p}
  \mathbf{\Sigma}_{\mathbf{q}_{\rm p}} = \alpha\kappa\left(p_{\rm p}\sum_{k=1}^Kd_k + 1\right)\mathbf{I}_{M\tau}\triangleq  \alpha\kappa\sigma_{\mathbf{q}_{\rm p}}^2.
\end{equation}

Substituting \eqref{C_g} \eqref{C_q_p} into \eqref{C_hat_g} and employing the matrix inverse lemma, we can easily obtain
\begin{align}\label{C_hat_g_1}
  \mathbf{\Sigma}_{\hat{\underline{\mathbf{g}}}} = \kappa^2\tau p_{\rm p}\left(\kappa^2\tau p_{\rm p}\mathbf{\Sigma}_{{\underline{\mathbf{g}}}}+(\kappa^2+\kappa\alpha\sigma_{\mathbf{q}_{\rm p}}^2)\mathbf{I}\right)^{-1}\mathbf{\Sigma}_{{\underline{\mathbf{g}}}}^2.
\end{align}

Thus the estimated vector $\hat{\mathbf{g}}_k$ of the $k$th user's channel $\mathbf{g}_k$ can be model as $\hat{\mathbf{g}}_k\sim\mathcal{CN}(\mathbf{0},\sigma_{\hat{\mathbf{g}}_k}^2\mathbf{I})$, where, according to \eqref{C_hat_g_1}, we have
\begin{equation}
  \sigma_{\hat{\mathbf{g}}_k}^2 = \frac{\kappa^2\tau p_{\rm p}d_k^2}{\kappa^2\tau p_{\rm p}d_k + (\kappa^2+\kappa\alpha\sigma_{\mathbf{q}_{\rm p}}^2)}.
\end{equation}

Similarly,the covariance matrix $\mathbf{\Sigma}_{\mathbf{e}_k}$ of channel estimate error $\mathbf{e}_k = \mathbf{g}_k - \hat{\mathbf{g}}_k$ can be expressed  as
\begin{equation}
  \mathbf{\Sigma}_{\mathbf{e}_k} = \E\{\mathbf{e}_k\mathbf{e}_k^H\} = (d_k-\sigma_{\hat{\mathbf{g}}_k^2})\mathbf{I} \triangleq \sigma_{\mathbf{e}_k}^2\mathbf{I}.
\end{equation}

According to \cite{kay1993fundamentals}, we can see that the channel estimate $\hat{\mathbf{g}}_k$ is independent of the estimation error $\mathbf{e}_k$ if the channel estimator in \eqref{hat_g} is used. In addition, we denote $\hat{\mathbf{G}} = \left[\hat{\mathbf{g}}_1,...,\hat{\mathbf{g}}_K\right]$ as the channel estimate matrix for all $K$ users.
\subsection{Uplink Transmission Stage}
In the  uplink transmission phase, we denote $\mathbf{R}^H$ as  the linear receiver. Then the quantized signal is processed as
\begin{equation}
  \hat{\mathbf{x}} = \mathbf{R}^H \mathbf{z}_{\rm u},
\end{equation}
and the $k$th element of $\hat{\mathbf{x}}$ is
\begin{align}\label{hat_x_k}
  \hat{x}_k & =   \kappa\sqrt{p_{\rm u}}\mathbf{r}_k^H\hat{\mathbf{g}}_k x_k  +\kappa\sqrt{p_{\rm u}}\mathbf{r}_k^H \hat{\mathbf{G}}_{(k)} \mathbf{x}_{(k)} \nonumber \\
  &~~+ \kappa\sqrt{p_{\rm u}}\mathbf{r}_k^H\mathbf{\mathcal{E}}\mathbf{x} + \kappa\mathbf{r}_k^H\mathbf{n}_{\rm u}+ \mathbf{r}_k^H\mathbf{q}_{\rm u} \nonumber \\
  & \triangleq \kappa\sqrt{p_{\rm u}}\mathbf{r}_k^H\hat{\mathbf{g}}_k x_k + \tilde{n}_k,
\end{align}
where $\mathbf{r}_k$ is the $k$th column of $\mathbf{R}$, $\mathbf{\mathcal{E}} = [\mathbf{e}_1,...,\mathbf{e}_K]$. $\hat{\mathbf{G}}_{(k)}$ is $\hat{\mathbf{G}}$ with the $k$th column removed and so as well as $\mathbf{x}_{(k)}$. We also refer the term $\tilde{n}_k$ as the effective noise.

Since $\hat{\mathbf{G}}$ and $\mathbf{\mathcal{E}}$ are independent for LMMSE channel estimation, the uplink SINR of the $k$th user is given by
\begin{equation}\label{SINR_k}
  {\textrm{SINR}}_k = \frac{\kappa^2{p_{\rm u}}|\mathbf{r}_k^H\hat{\mathbf{g}}_k|^2}{\kappa^2{p_{\rm u}}\|\mathbf{r}_k^H \hat{\mathbf{G}}_{(k)}\|_2^2 + \kappa^2 p_{\rm u}\mathbf{r}_k^H \mathbf{\Sigma}_{\mathbf{\mathcal{E}}} \mathbf{r}_k + \textrm{AQN}_k},
\end{equation}
where
\begin{equation}
  \mathbf{\Sigma}_{\mathbf{\mathcal{E}}} = \E\{\mathbf{\mathcal{E}}\mathbf{\mathcal{E}}^H\} = \sum_{k=1}^K\sigma_{\mathbf{e}_k}^2 \mathbf{I},
\end{equation}
\begin{equation}
  \textrm{AQN}_k = \kappa^2\mathbf{r}_k^H\mathbf{r}_k+\mathbf{r}_k^H\mathbf{\Sigma}_{\mathbf{q}_{\rm u}}\mathbf{r}_k.
\end{equation}

It is shown in \cite{hassibi2003how} that for unquantized system, the MMSE channel estimator is the only estimator renders the additive channel estimation error uncorrelated with the data signals. For the quantized system considered in this paper, we can also readily obtain the correlation between the additive noise due to the estimation error  and the data symbols based on the orthogonality property of MMSE as
\begin{equation}
  \E\{\mathbf{\mathcal{E}}\mathbf{x}\mathbf{x}^H|\mathbf{G}\} = \E\{\mathbf{\mathcal{E}}|\mathbf{G}\}\E\{\mathbf{x}\mathbf{x}^H|\mathbf{G}\} = \mathbf{0}.
\end{equation}
This implies that in the low-resolution systems, the channel estimation error obtained by the LMMSE channel estimator (and hence the effective noise $\tilde{n}_k$) are uncorrelated with the desired signal. Therefore, based on the fact that the additive Gaussian noise minimizes the mutual information between the input and output, a lower bound on the achievable rate of the $k$th user can be given by
\begin{equation}\label{rate_gener}
R_k=\log_2\left(1+\textrm{SINR}_k\right).
\end{equation}

\section{Asymptotic Achievable Rate Analysis with LMMSE Receiver}

%
%
In this section, we will focus on deriving the achievable rate with MMSE receiver, which is properly designed by taking not only the AWGN noise, but also the channel estimation error  and the quantizer noise into account, for low-resolution  massive MIMO system. We also assume that  both  the  number  of  users $K$ and  BS antennas $M$ approach infinity with their ratio $\xi = K/M\rightarrow\infty$ is  bounded.  As can be seen later, the  Stieltjes  transform  of  the large dimension channel matrix $\mathbf{G}$ is employed in order to obtain the asymptotic achievable rate.

According to \cite{kay1993fundamentals}, the MMSE receiver of the $k$th user for a given channel estimate $\hat{\mathbf{G}}$ can be expressed as
\begin{equation}\label{MMSE_receiver}
  \mathbf{r}_k^H = \hat{\mathbf{g}}_k^H\left(\hat{\mathbf{G}}\hat{\mathbf{G}}^H + \mathbf{\Xi}　\right)^{-1},
\end{equation}
where
\begin{equation}
  \mathbf{\Xi} = \mathbf{\Sigma}_{\mathbf{\mathcal{E}}} + \frac{1}{\kappa^2p_{\rm u}}\mathbf{\Sigma}_{\mathbf{q}_{\rm u}} + \frac{1}{p_{\rm u}}\mathbf{I}.
\end{equation}

It should be noted that, in contrast to the typical MMSE receivers, the receiver in \eqref{MMSE_receiver} takes not only the AWGN but also the channel estimation error and the quantization noise into account. As we will see later, this improves the system performance.

In order to derive an asymptotic equivalent expression of $R_k$, we first provide an approximation of the covariance matrix $\mathbf{\Sigma}_{\mathbf{q}_{\rm u}}$. We note that, based on the AQNM model, the covariance matrix $\mathbf{\Sigma}_{\mathbf{q}_{\rm u}}$ is a diagonal matrix, which depends on the individual channel realization. For massive MIMO systems with $K\rightarrow\infty$, we can approximate $\mathbf{\Sigma}_{\mathbf{q}_{\rm u}}$ as
\begin{equation}\label{C_q_u_app}
  \mathbf{\Sigma}_{\mathbf{q}_{\rm u}} \cong \alpha \kappa \left( p_{\rm u}\sum_{k=1}^Kd_k + 1\right)\mathbf{I} \triangleq \alpha \kappa\sigma_{\mathbf{q}_{\rm u}}^2\mathbf{I}.
\end{equation}
This is due to the approximation that $\diag\left(\mathbf{G}\mathbf{G}^H\right)\cong \sum_{k=1}^Kd_k\mathbf{I}$ when $K\rightarrow\infty$. Thus, in the massive MIMO configuration, the MMSE receiver in \eqref{MMSE_receiver} for the $k$th user can be approximated as
\begin{equation}\label{MMSE_receiver_app}
  \mathbf{r}_k^H = \hat{\mathbf{g}}_k^H\left(\hat{\mathbf{G}}\hat{\mathbf{G}}^H + \theta\mathbf{I}\right)^{-1} \triangleq \hat{\mathbf{g}}_k^H\mathbf{A}^{-1},
\end{equation}
where $\theta = \sum_{k=1}^K \sigma_{\mathbf{e}_k}^2 + (\kappa+\alpha\sigma_{\mathbf{q}_{\rm u}}^2)/\kappa p_{\rm u}$.

Accordingly, the $k$th user's achievable rate  in \eqref{rate_gener} can be approximated by
\begin{equation}\label{R_k_gener}
  R_k=\log_2\left(1+\frac{|\hat{\mathbf{g}}_k^H\mathbf{A}^{-1}\hat{\mathbf{g}}_k|^2}{\|\hat{\mathbf{g}}_k^H\mathbf{A}^{-1} \hat{\mathbf{G}}_{(k)}\|^2_2 + \theta \hat{\mathbf{g}}_k^H\mathbf{A}^{-2}\hat{\mathbf{g}}_k }\right).
\end{equation}

With the approximation above, we can obtain  the asymptotic equivalent of $R_k$ shown in Theorem 1.

\begin{figure*}[!t]
\normalsize
\setcounter{MYtempeqncnt}{\value{equation}}
\setcounter{equation}{27}
\begin{equation}\label{theorem_1_eq}
R_k = \log_2\left(1+\frac{\sigma_{\hat{\mathbf{g}}_k}^4 M K (\theta+\psi)^2(\lambda-1)}{(\lambda\theta+\psi(\lambda-1))(\sigma_{\hat{\mathbf{g}}_k}^4 M^2\theta+\theta(\theta+\psi)^2+\sigma_{\hat{\mathbf{g}}_k}^2(\theta+\psi)(K\theta+K\psi+2M\theta))}\right)
\end{equation}
\setcounter{equation}{28}
\hrulefill
\vspace*{-0.3cm}
\end{figure*}

{\it Theorem 1:} If the LMMSE channel estimator in \eqref{hat_g} and the proper designed MMSE receiver in \eqref{MMSE_receiver} are used in the low-resolution massive MIMO systems, then the asymptotic equivalent for the $k$th user's achievable rate can be expressed as \eqref{theorem_1_eq} shown on the next page, where $\psi$ and $\lambda$ can be calculated as \eqref{38} and \eqref{40}, respectively.
\begin{IEEEproof}
We first focus on deriving the asymptotic equivalent of the desired signal term $\hat{\mathbf{g}}_k^H\mathbf{A}^{-1}\mathbf{g}_k$. By observing that $\left(\hat{\mathbf{G}}\hat{\mathbf{G}}^H+\theta\mathbf{I}\right)^{-1} = \left(\hat{\mathbf{G}}_{(k)}\hat{\mathbf{G}}_{(k)}^H + \hat{\mathbf{g}}_k\hat{\mathbf{g}}_k^H+\theta\mathbf{I}\right)^{-1}$ and applying matrix inverse lemma, we have
\begin{equation}\label{aaaa}
  \hat{\mathbf{g}}_k^H\mathbf{A}^{-1}\hat{\mathbf{g}}_k = \frac{\hat{\mathbf{g}}_k^H\mathbf{A}_{(k)}^{-1}\hat{\mathbf{g}}_k}{1+\hat{\mathbf{g}}_k^H\mathbf{A}_{(k)}^{-1}\hat{\mathbf{g}}_k},
\end{equation}
where $\mathbf{A}_{(k)} = \hat{\mathbf{G}}_{(k)}\hat{\mathbf{G}}_{(k)}^H +\theta\mathbf{I}$. Since $\hat{\mathbf{g}}_k$ is independent of $\mathbf{A}_{(k)}$ and has variance of $\sigma_{\hat{\mathbf{g}}_k}^2$, we can sequentially employ the trace lemma and rank-1 perturbation to obtain
\begin{align}\label{30}
  \hat{\mathbf{g}}_k^H\mathbf{A}_{(k)}^{-1}\hat{\mathbf{g}}_k \xrightarrow{a.s.}\sigma_{\hat{\mathbf{g}}_k}^2\tr\left(\mathbf{A}^{-1}\right).
\end{align}

 Thus the asymptotic equivalent of  the  desired signal power is given by
\begin{equation}
  |\hat{\mathbf{g}}_k^H\mathbf{A}^{-1}\hat{\mathbf{g}}_k|^2 \xrightarrow{a.s.} \frac{\sigma_{\hat{\mathbf{g}}_k}^4\tr\left(\mathbf{A}^{-1}\right)^2}{(1+\sigma_{\hat{\mathbf{g}}_k}^2\tr\left(\mathbf{A}^{-1}\right))^2}.
\end{equation}

Then we deal with the asymptotic equivalent of the user interference term $\|\hat{\mathbf{g}}_k^H\mathbf{A}^{-1} \hat{\mathbf{G}}_{(k)}\|^2_2$.  In order to eliminate the dependence between $\hat{\mathbf{g}}_k$ and $\mathbf{A}$, we rewrite $\|\hat{\mathbf{g}}_k^H\mathbf{A}^{-1} \hat{\mathbf{G}}_{(k)}\|^2_2$ as
\begin{align}\label{Interf}
  &\|\hat{\mathbf{g}}_k^H\mathbf{A}^{-1} \hat{\mathbf{G}}_{(k)}\|^2_2 \nonumber \\
   &= \hat{\mathbf{g}}_k^H\left(\mathbf{A}^{-1}_{(k)}+\mathbf{A}^{-1}-\mathbf{A}^{-1}_{(k)}\right) \hat{\mathbf{G}}_{(k)}\hat{\mathbf{G}}_{(k)}^H\mathbf{A}^{-1}\hat{\mathbf{g}}_k \nonumber \\
  & = \hat{\mathbf{g}}_k^H\mathbf{A}^{-1}_{(k)}\hat{\mathbf{G}}_{(k)}\hat{\mathbf{G}}_{(k)}^H\mathbf{A}^{-1}\hat{\mathbf{g}}_k \nonumber \\
   &~~~+ \hat{\mathbf{g}}_k^H\left(\mathbf{A}^{-1}-\mathbf{A}^{-1}_{(k)}\right) \hat{\mathbf{G}}_{(k)}\hat{\mathbf{G}}_{(k)}^H\mathbf{A}^{-1}\hat{\mathbf{g}}_k.
\end{align}

By noting that $\mathbf{A}^{-1} - \mathbf{A}_{(k)}^{-1} = -\mathbf{A}^{-1}\hat{\mathbf{g}}_{k}\hat{\mathbf{g}}_k^H \mathbf{A}^{-1}_{(k)}$ and $\mathbf{A}_{(k)}^{-1}\hat{\mathbf{G}}_{(k)}\hat{\mathbf{G}}_{(k)}^H = \mathbf{I} - \theta\mathbf{A}_{(k)}^{-1}$, \eqref{Interf} can be further simplified as
\begin{align}
  &\|\hat{\mathbf{g}}_k^H\mathbf{A}^{-1} \hat{\mathbf{G}}_{(k)}\|^2_2 = \hat{\mathbf{g}}_k^H\mathbf{A}^{-1}\hat{\mathbf{g}}_k - \theta\hat{\mathbf{g}}_k^H\mathbf{A}^{-1}_{(k)}\mathbf{A}^{-1}\hat{\mathbf{g}}_k \nonumber \\
  &~~~-\hat{\mathbf{g}}_k^H\mathbf{A}^{-1}\hat{\mathbf{g}}_k(\hat{\mathbf{g}}_k^H\mathbf{A}^{-1}\hat{\mathbf{g}}_k - \theta\hat{\mathbf{g}}_k^H\mathbf{A}^{-1}_{(k)}\mathbf{A}^{-1}\hat{\mathbf{g}}_k).
\end{align}

According to Lemma 7 in \cite{wagner2012large}, we can obtain the asymptotic equivalent of $\hat{\mathbf{g}}_k^H\mathbf{A}^{-1}_{(k)}\mathbf{A}^{-1}\hat{\mathbf{g}}_k$ as
\begin{align}\label{asda}
  \hat{\mathbf{g}}_k^H\mathbf{A}^{-1}_{(k)}\mathbf{A}^{-1}\hat{\mathbf{g}}_k \xrightarrow{a.s.} \frac{\sigma_{\hat{\mathbf{g}}_k}^2\tr(\mathbf{A}^{-2}_{(k)})}{\sigma_{\hat{\mathbf{g}}_k}^2\tr(\mathbf{A}^{-1}_{(k)})+1} \xrightarrow{a.s.}\frac{\sigma_{\hat{\mathbf{g}}_k}^2\tr(\mathbf{A}^{-2})}{\sigma_{\hat{\mathbf{g}}_k}^2\tr(\mathbf{A}^{-1})+1}.
\end{align}
where the second step in \eqref{asda} is based on the rank-1 perturbation lemma. Therefore, combining with \eqref{aaaa} we can further obtain the asymptotic equivalent of $\|\hat{\mathbf{g}}_k^H\mathbf{A}^{-1} \hat{\mathbf{G}}_{(k)}\|^2_2$ as
\begin{equation}\label{37}
  \|\hat{\mathbf{g}}_k^H\mathbf{A}^{-1} \hat{\mathbf{G}}_{(k)}\|^2_2 \xrightarrow{a.s.} \frac{\sigma_{\hat{\mathbf{g}}_k}^2(\tr(\mathbf{A}^{-1})-\theta\tr(\mathbf{A}^{-2}))}{(1+\sigma_{\hat{\mathbf{g}}_k}^2\tr(\mathbf{A}^{-1}))^2}.
\end{equation}

Next we focus on the quantization noise and white noise term $\hat{\mathbf{g}}_k^H\mathbf{A}^{-2}\hat{\mathbf{g}}_k$. Note that in the massive MIMO system with large $M$ and $K$, we can approximate $\hat{\mathbf{g}}_k^H\mathbf{A}^{-2}\hat{\mathbf{g}}_k$ as \begin{align}\label{ccc}
  \hat{\mathbf{g}}_k^H\mathbf{A}^{-2}\hat{\mathbf{g}}_k &\cong \frac{1}{K} \tr(\hat{\mathbf{G}}^H\mathbf{A}^{-2}\hat{\mathbf{G}}) = \frac{1}{K}\tr(\hat{\mathbf{G}}\hat{\mathbf{G}}^H\mathbf{A}^{-2}) \nonumber \\
  & =\frac{1}{K}(\tr(\mathbf{A}^{-1}) - \theta\tr( \mathbf{A}^{-2})).
\end{align}
The last step in \eqref{ccc} is due to $\hat{\mathbf{G}}\hat{\mathbf{G}}^H\mathbf{A}^{-2}= \mathbf{A}^{-1} - \theta \mathbf{A}^{-2}$.

We can see from \eqref{30}, \eqref{37} and \eqref{ccc} that the $\textrm{SINR}_k$ is related to the terms of $\tr(\mathbf{A}^{-1})$ and $\tr(\mathbf{A}^{-2})$. Fortunately, $\tr(\mathbf{A}^{-2})$ is the derivative of $-\tr(\mathbf{A}^{-1})$ and the term of $\tr(\mathbf{A}^{-1})$ can be obtained by using the Stieltjes transform of $\hat{\mathbf{G}}\hat{\mathbf{G}}^H$. More precisely, by using the Stieltjes transform of matrix $\hat{\mathbf{G}}\hat{\mathbf{G}}^H$, we can further obtain
\begin{equation}
  \frac{1}{M}\tr\left(\mathbf{A}^{-1}\right)\xrightarrow{a.s.} \frac{1}{\theta+\psi},
\end{equation}
where $\psi$ is the unique real positive solution of
\begin{equation}\label{38}
  \psi = \tr\left(\hat{\mathbf{D}}\left(\mathbf{I} + \frac{M}{\theta+\psi}\hat{\mathbf{D}}\right)^{-1}\right),
\end{equation}
with $\hat{\mathbf{D}} = \diag\left(\sigma_{\hat{\mathbf{g}}_1}^2,...,\sigma_{\hat{\mathbf{g}}_K}^2\right)$. And
\begin{equation}
  \frac{1}{M}\tr\left(\mathbf{A}^{-2}\right) = -\frac{1}{M}\frac{\partial\tr\left(\mathbf{A}^{-1}\right)}{\partial \theta}\xrightarrow{a.s.} \frac{1}{1-\lambda}\frac{1}{(\theta+\psi)^2},
\end{equation}
where
\begin{equation}\label{40}
  \lambda = -\frac{M}{(\theta+\psi)^2}\tr\left(\hat{\mathbf{D}}^2\left(\mathbf{I} + \frac{M}{\theta+\psi}\hat{\mathbf{D}}\right)^{-2}\right).
\end{equation}

Combining all the asymptotic equivalents obtain above, we can finally arrive Theorem 1.
\end{IEEEproof}

Note that the asymptotic equivalent for the achievable rate in \eqref{theorem_1_eq} includes implicit equation, which is difficult to obtain insights into the performance for the low-resolution massive MIMO systems. In order to simplify the analysis and obtain insights into the performance for the low-resolution massive MIMO systems, we consider the special case of ignoring the large scale fading coefficients and provide the asymptotic equivalent  for the achievable rate as in Proposition 1.

{\it Proposition 1:} For the special case of $d_1=...=d_K= 1$, the asymptotic equivalent expression in \eqref{theorem_1_eq} can be reduced as
\begin{equation}\label{Prop1}
  \tilde{R}_k = \log_2\left(1+\frac{MK\sigma^4 a^2}{(a-\theta b)(K\sigma^2+\theta(1+\sigma^2 M a)^2)}\right).
\end{equation}
\begin{IEEEproof}
  In the special case of $d_1=...=d_K=1$, the Stieltjes transform of matrix $\hat{\mathbf{G}}\hat{\mathbf{G}}^H$ at the point $-\theta$ converges to \eqref{u} shown on the next page almost surely  \cite[Example 2.8]{tulino2004random}. Thus, by calculating the derivative of $\tr(\mathbf{A}^{-1})$ along $\theta$, we can obtain $\tr(\mathbf{A}^{-2})$ as shown in \eqref{v} on the next page.

  Therefore, in this case of $d_1=...=d_K=1$, we derive the asymptotic equivalent of the achievable rate as in Proposition~1.
\end{IEEEproof}

\begin{figure*}[!t]
\normalsize
\setcounter{MYtempeqncnt}{\value{equation}}
\setcounter{equation}{41}
\begin{align}
\label{u}\frac{1}{M}\tr(\mathbf{A}^{-1}) &\xrightarrow{a.s.} \frac{1}{2}\left(\sqrt{\frac{(1-\xi)^2}{\theta^2}+\frac{2(1+\xi)}{M\sigma^2\theta}+\frac{1}{M^2\sigma^4}}+\frac{(1-\xi)}{\theta}-\frac{1}{M\sigma^2}\right) \triangleq a, \\
\label{v}\frac{1}{M}\tr(\mathbf{A}^{-2}) &\xrightarrow{a.s.} \frac{1}{2}\left(\frac{1}{2}\left(\frac{(1-\xi)^2}{\theta^2}+\frac{2(1+\xi)}{M\sigma^2\theta}+\frac{1}{M^2\sigma^4}\right)^{-\frac{1}{2}}
    \left(\frac{2(1-\xi)^2}{\theta^3} + \frac{2(1+\xi)}{M\sigma^2\theta^2}\right)+\frac{(1-\xi)}{\theta^2}\right)\triangleq b.
\end{align}
\setcounter{equation}{43}
\hrulefill
\vspace*{-0.3cm}
\end{figure*}

\subsection{Performance Evaluation}
In this subsection, we will employ the asymptotic equivalent expression as in Proposition 1 to obtain the insights into the performance of the low-resolution massive MIMO system.

{\it Remark 1:} With fixed $p_{\rm p}$, $p_{\rm u}$, $M$ and $K$, when the quantization resolution increase to infinity, then \eqref{Prop1} reduces to
\begin{equation}
\footnotesize
  \tilde{R}_k \rightarrow \log_2\left(1+\frac{MK\sigma_{1}^4 a_{1}^2}{(u_{1}-\theta b_{1})(K\sigma_{1}^2+\theta_{1}(1+\sigma_{1}^2 M a_{1})^2)}\right),
\end{equation}
where $\sigma_{1}^2$ and $\theta_{1}$ are given by
\begin{align}
  \sigma_{1}^2 &= \frac{\tau p_{\rm p}}{\tau p_{\rm p} + 1}, \\
  \theta_{1} &= \frac{K}{\tau p_{\rm p} + 1} + \frac{1}{p_{\rm u}},
\end{align}
respectively. $a_{1}$ and $b_{1}$ are the same with \eqref{u} and \eqref{v}, but replacing the parameters $\sigma$ and $\theta$ with $\sigma_{1}$ and $\theta_{1}$, respectively.

This conclusion is reasonable because in the systems with the infinity resolution ADCs, the quantizer noise can be ignored (i.e., $\kappa\rightarrow 1$). We also notice that the quantization process affects both the desired signal term and the effective noise term since the parameters $a$, $b$ and $\sigma^2$ contains the quantization factor $\kappa$. This implies that the quantization noise cannot be simply modeled as an additive noise that independent with the quantizer input signal as shown in \cite{emil2014massive}.

{\it Remark 2:} With fixed $M$, $K$ and quantization resolution, when the transmit power $p_{\rm u}$ and $p_{\rm p}$ both increase to infinity, then \eqref{Prop1} converges to
\begin{equation}\label{Prop2}
  \tilde{R}_k \rightarrow \log_2\left(1+\frac{MK\sigma_{2}^4 u_{2}^2}{(a_{2}-\theta b_{2})(K\sigma_{2}^2+\theta_{2}(1+\sigma_{2}^2 M a_{2})^2)}\right),
\end{equation}
where $\sigma_{2}^2$ and $\theta_{2}$ are given by
\begin{align}
  \label{sss}\sigma_{2}^2 &= \frac{\kappa\tau}{\kappa\tau+K\alpha} \\
  \theta_{2} &= \frac{K^2\alpha}{\kappa\tau+K\alpha} + \frac{K\alpha}{\kappa},
\end{align}
respectively. $a_{2}$ and $b_{2}$ are the same with \eqref{u} and \eqref{v}, but replacing the parameters $\sigma$ and $\theta$ with $\sigma_{2}$ and $\theta_{2}$.

As we can see from \eqref{Prop2} that the achievable rate is fixed when the transmit power increase to infinity and depends on the quantization factor $\kappa$. This implies that, in contrast to the traditional massive MIMO systems with infinite quantization resolution, the performance loss caused by the low-resolution ADCs cannot be compensated for by simply increasing the transmit power. Moreover, \eqref{sss} reveals that when the pilot transmit power increase to infinity, the MSE performance of the LMMSE channel estimator in \eqref{hat_g} converges to a fixed point. That is to say, in contrast to the traditional systems with infinite resolution ADCs, perfect CSI cannot be acquired by merely increasing the pilot transmit power in the low-resolution systems.

Next we will study how to compensate the performance loss due to the low-resolution ADCs. As it is known that for traditional massive MIMO systems, the achievable rate achieved by MMSE receiver will increase to infinity with increasing the transmit power, while the achievable rate saturates to a fixed point with infinite transmit power for low-resolution massive MIMO systems. Therefore, we will investigate the problem of how many more antennas would be needed in low-resolution massive MIMO systems to achieve the same achievable rate as that in traditional massive MIMO systems.

We assume that the number of antennas in low-resolution massive MIMO and traditional massive MIMO systems are $M_{\rm low}$ and $M_{\rm conv}$, respectively. We also assume that, in both scenarios, the number of users $K$ is fixed, the transmit power of users is $p$, such that $p_{\rm p} = p_{\rm u} = p$ and the training length $\tau =K$. We aim to seek for the minimum value of $M_{\rm low}$ such that both quantized and unquantized systems achieve the same achievable rate and provide the following optimization problem:
\begin{align}\label{xxxxx}
  &{\rm minimize}&& M_{\rm low} \nonumber\\
  &{\rm subject~to}&&\tilde{R} = \tilde{R}_{\rm conv},
\end{align}
where $\tilde{R}_{\rm conv} $ is given by \eqref{Prop1}. Although  a simple closed-form of the optimal $M_{\rm low}$ cannot be acquired, we can numerically determine the the optimal $M_{\rm low}$ since the optimization problem in \eqref{xxxxx} only contains one simple variable. 

\section{Numerical Results}
In this section,  a low-resolution massive MIMO systems with $K = 50$ and $\tau = K$ is considered. The large scale coefficients for all users are assume to be equal, i.e., $d_{1}= ...= d_K = 1$. 

\begin{figure}
  \centering
  \includegraphics[width=9cm]{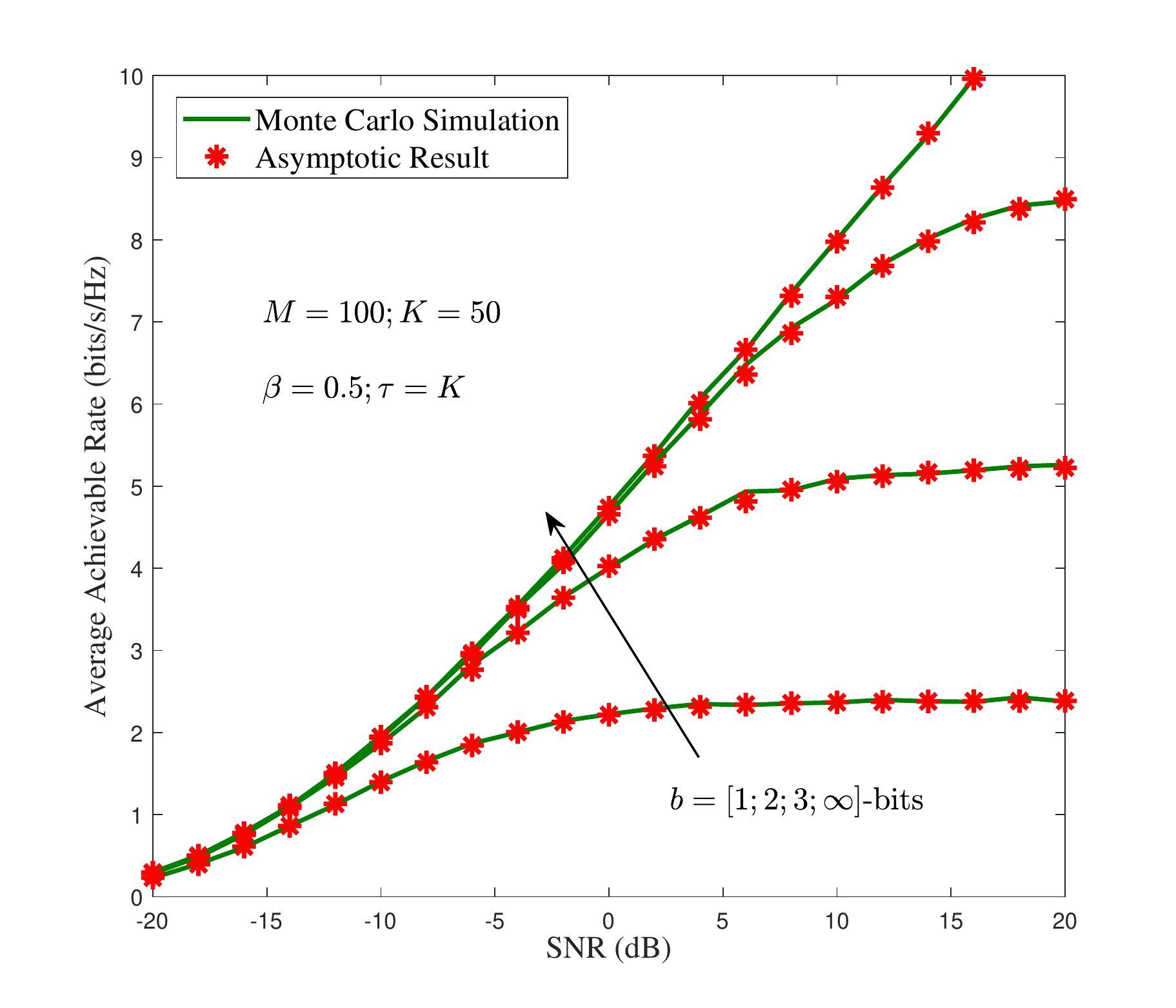}\\
  \caption{Average achievable rate versus the SNR with the number of antennas $M =100$ for different quantizers with 1, 2, 3 and $\infty$ bits.}\label{fig1}
\end{figure}
We first evaluate the validity of the asymptotic equivalent for the achievable rate given in Proposition 1. Fig.~\ref{fig1} shows the average achievable rate versus the SNR with the number of antennas $M =100$ for four different quantizer bits with 1, 2, 3 and $\infty$ bits. The solid lines represent the achievable rate plotted by  using Monte-Carlo method, and the markers are calculated by using the asymptotic expression given in \eqref{Prop1}. It can be easily see from Fig.~\ref{fig1} that the results obtained by our analysis almost overlaps with the Monte-Carlo simulations. This verifies the accuracy of our obtained result in \eqref{Prop1}. 

\begin{figure}
  \centering
  \includegraphics[width=9cm]{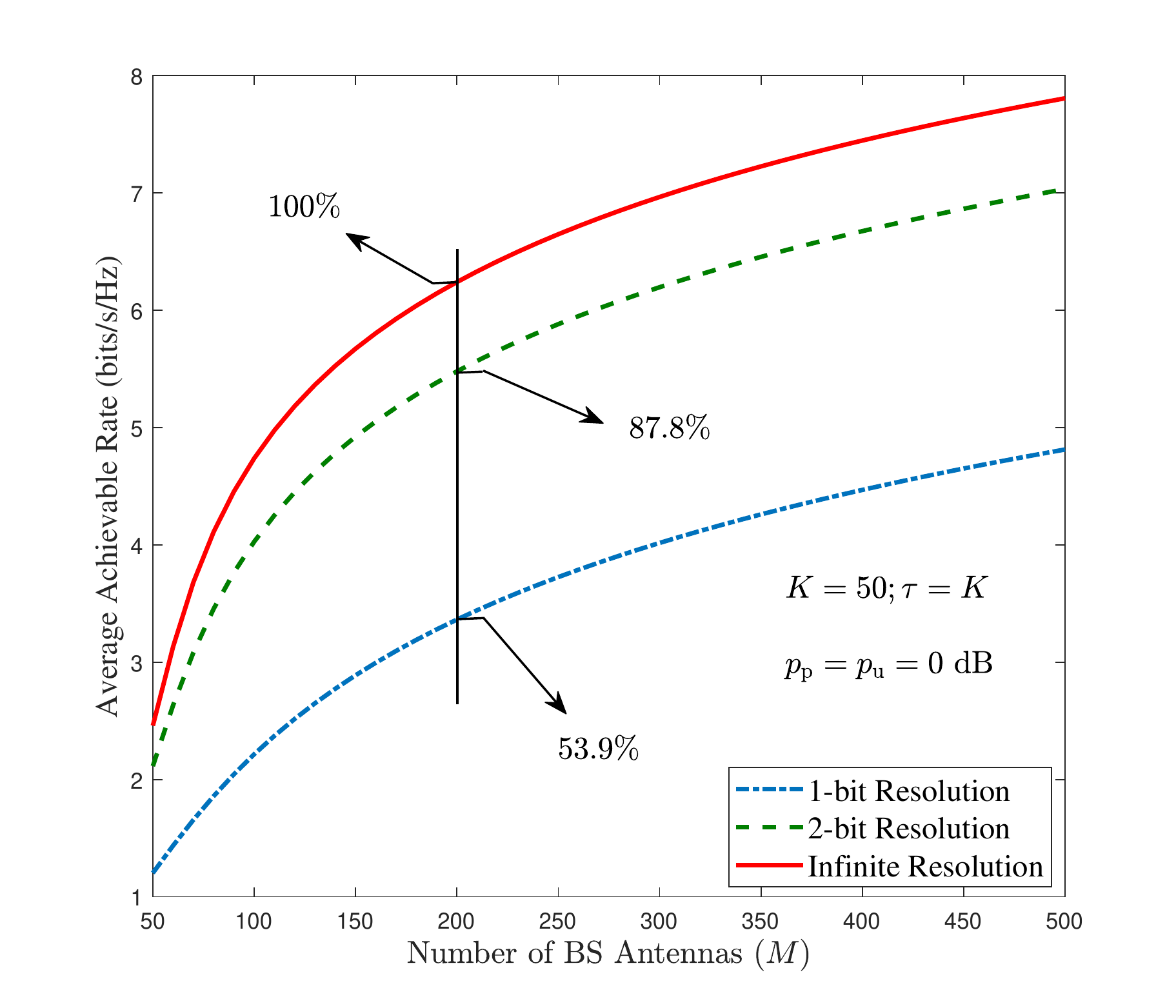}\\
  \caption{Average achievable rate versus number of receive
antennas for different quantizers with 1, 2 and $\infty$ bits with average transmit power $p_{\rm p}=p_{\rm u} = 0$dB.}\label{fig2}
\end{figure}

We then compare the average achievable rate  between the low-resolution and traditional massive MIMO systems. In Fig.~\ref{fig2}, the average achievable rates  versus the number of receive antennas for with transmit power $p_{\rm p}=p_{\rm u} = 0$dB is illustrated. We can see from Fig.~\ref{fig2} that, compared with the traditional system, the low-resolution massive MIMO systems can still achieve a reasonable high average achievable rate. For example, with $M=200$, the system with 1 quantization bits and 2 quantization bits can still achieve an average rate of 3.36 bits/s/Hz and 5.48 bits/s/Hz, respectively, which  amounts to $54.9\%$ and $87.8\%$ of the average achievable rate of the traditional
system. We can also see from Fig.~\ref{fig2} that we can increase the number of antennas at the base station to compensate for the performance loss due to the low-resolution ADCs.

\begin{figure}
  \centering
  \includegraphics[width=9cm]{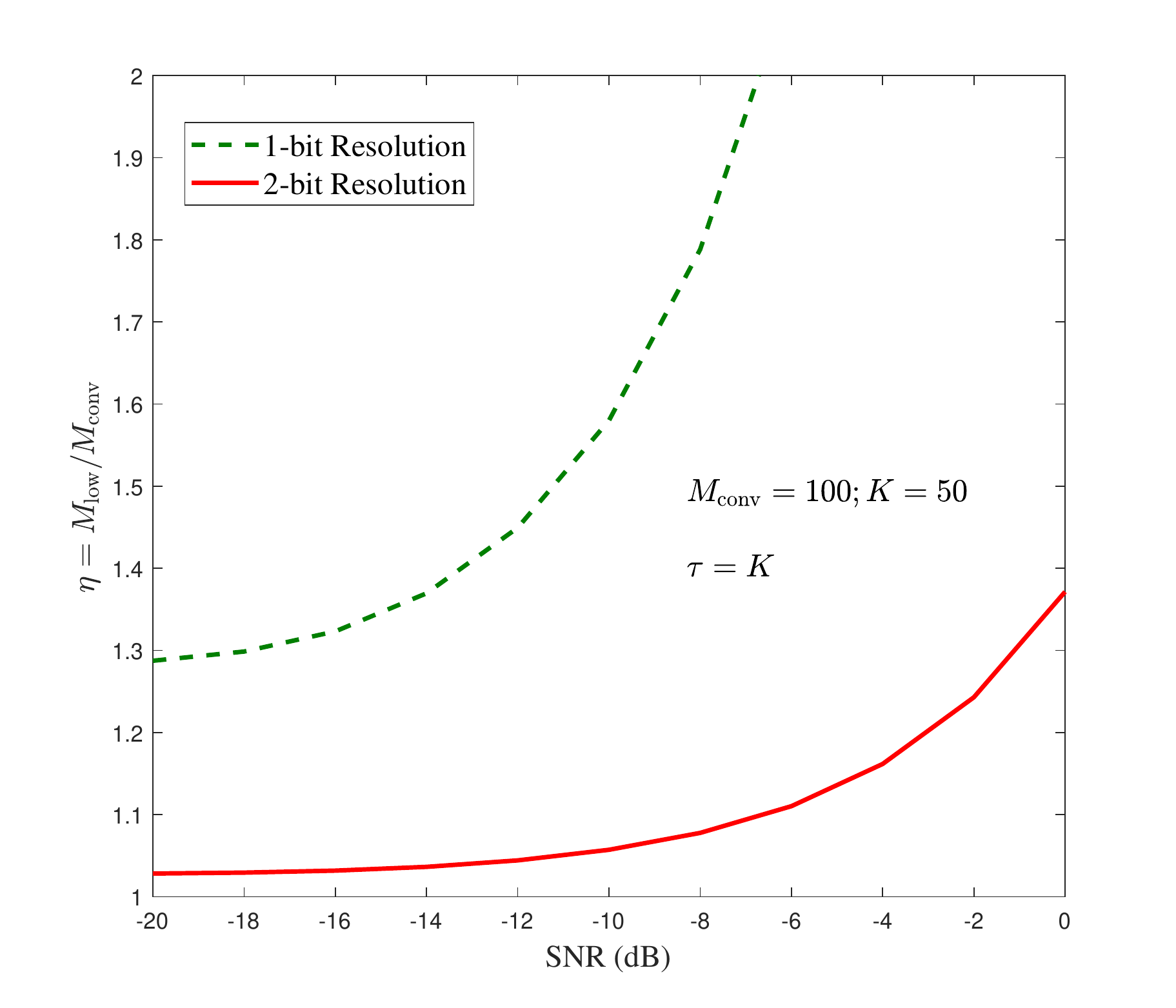}\\
  \caption{The ratio of $\eta= M_{\rm low}/M_{\rm conv}$ versus SNR with $M_{\rm conv}=100$ and $K = 50$ for different quantizer bits.}\label{fig3}
\end{figure}

In order to elaborate how to compensate the performance loss in low-resolution massive MIMO systems by increasing the number of antennas, Fig.~\ref{fig3} illustrates the ratio of $\eta = M_{\rm low}/M_{\rm conv}$ when both low-resolution and traditional systems achieve the same achievable rate with the assumptions of  $M_{\rm conv} = 100$ and  $K = 50$. We can see that in the low SNR region (i.e., ${\rm SNR}<-10$dB), the systems with 1-bit quantization needs to equip around 1.5 times more antennas to achieve the same average achievable rate as the traditional systems, while for the systems with 2-bit quantization bits, the ratio decrease to 1.1. This is because that with  higher quantization bits, the  performance loss is less. Moreover, as the SNR grows large, the ratio also goes to infinity. This is because that the achievable rate of the traditional systems increase to infinity with the increasing SNR. However, as shown in Remark~2, the achievable rate of the low-resolution systems saturates to a fixed point even if the SNR increases to infinity.

\section{Conclusions}

The uplink achievable rate performance of the low-resolution massive MIMO systems is considered in this paper. By using the AQNM, we derived a tight asymptotic equivalent for the uplink achievable rate with a proper designed MMSE receiver assuming imperfect CSI is known at the BS. It is shown  that the performance loss due to the low-resolution quantization is not as severe as might be imagined. Numerical results demonstrated that for 2-bit quantization systems, only 1.1 times more antennas are needed to be installed at the BS for low-resolution massive MIMO systems such that the   performance loss can be compensated. 
\bibliographystyle{IEEEtran}
\bibliography{reference}

\end{document}